\documentstyle[12pt]{article}

\makeatletter

\def\section{\@startsection {section}{1}{\z@}{-3.5ex plus -1ex minus
 -.2ex}{2.3ex plus .2ex}{\large\bf}}
\def\subsection{\@startsection{subsection}{2}{\z@}{-3.25ex plus -1ex minus
 -.2ex}{1.5ex plus .2ex}{\normalsize\bf}}
\makeatother

\begin{document}
\global\parskip=4pt


\newcommand{\nc}{\newcommand}
\newcommand{\rnc}{\renewcommand}

\makeatletter
\@addtoreset{equation}{section}
\rnc{\theequation}{\arabic{section}.\arabic{equation}}
\makeatother

\nc{\ignore}[1]{}

\nc{\be}{\begin{equation}}
\nc{\ee}{\end{equation}}
\nc{\bea}{\begin{eqnarray}}
\nc{\eea}{\end{eqnarray}}

\rnc{\a}{\alpha}
\rnc{\b}{\beta}
\rnc{\d}{\delta}
\nc{\e}{\eta}
\nc{\eb}{\bar{\eta}}
\nc{\f}{\phi}
\nc{\fb}{\bar{\phi}}
\nc{\vf}{\varphi}
\nc{\p}{\psi}
\rnc{\pb}{\bar{\psi}}
\rnc{\c}{\chi}
\nc{\cb}{\bar{\c}}
\rnc{\l}{\lambda}
\rnc{\o}{\omega}
\rnc{\t}{\theta}
\nc{\T}{D_0}
\nc{\bt}{}
\nc{\tb}{\bar{\theta}}
\nc{\ep}{\epsilon}
\nc{\im}{\imath}
\nc{\tc}{\tilde{C}}
\nc{\tB}{\tilde{B}}
\nc{\tJ}{\tilde{J}}



\nc{\trac}[2]{{\textstyle\frac{#1}{#2}}}

\nc{\mat}[4]{\left(\begin{array}{cc}#1&#2\\#3&#4\end{array}\right)}

\def\tr{\mathop{\rm tr}\nolimits}
\def\Tr{\mathop{\rm Tr}\nolimits}
\nc{\r}{\rightarrow}
\nc{\ot}{\otimes}
\rnc{\ss}{\subset}
\nc{\ra}{\,\rangle}
\nc{\la}{\langle\,}
\nc{\ad}{a^\dagger}
\nc{\bd}{b^\dagger}
\nc{\rf}[1]{(\ref{#1})}
\rnc{\ll}{\label}
\nc{\del}{\partial}
\rnc{\=}{ \; = \; }
\rnc{\-}{ \: - \: }
\rnc{\+}{ \: + \: }
\nc{\bs}{ B^s_+ \, B^s_- }
\nc{\bw}{ B^w_+ \, B^w_- }
\rnc{\sb}{ B^s_- \, B^s_+ }
\nc{\wb}{ B^w_- \, B^w_+ }
\nc{\qnb}{\{ \, n_b \, \} }
\nc{\eq}[1]{(\ref{#1})}
\nc{\com}[3]{[ \: #1 \: ,\: #2 \: ] &=& #3}
\nc{\tcom}[3]{[ \: #1 \: ,\: #2 \: ] \= #3}

\nc{\CV}{Calogero-Vasiliev}
\nc{\os}{oscillator}
\nc{\ie}{{\em i.e.}}
\nc{\HW}{Heisenberg-Weyl}
\nc{\HWA}{Heisenberg-Weyl algebra}

\begin{titlepage}
\vskip 1.in
\begin{center}
{\LARGE\bf Deformed Algebras from \\Inverse Schwinger Method}\\
\vskip .25in
\vskip .5in
{\bf Kyung-Hyun Cho} 
\vskip .10in
Department of Physics\\
Chonbuk National University\\
Chonju, \hspace{.1cm} {\bf 560-756},\hspace{.2cm} {\bf Korea}
\vskip .5in
{\bf S. U. Park} 
\vskip .10in
Department of Physics\\
Jeonju University\\
Chonju, \hspace{.1cm} {\bf 560-759},\hspace{.2cm} {\bf Korea}
\vskip .15in
\end{center}
\vskip 1.in
\begin{abstract}
We consider a problem which may be viewed as
an inverse one to the Schwinger realization of Lie algebra,
and suggest a procedure of deforming the so-obtained algebra.
We illustrate the method through a few simple examples
extending Schwinger's $su(1,1)$ construction.
As results, various q-deformed algebras are (re-)produced as well as
their undeformed counterparts. Some extensions of the method are
pointed out briefly.
\end{abstract}
\end{titlepage}


\section{Introduction}

The Schwinger method of realizing $su(2)$ algebra in terms of the products of
two independent sets of Heisenberg-Weyl (or \os, boson) algebras\cite{sch}
is a typical example of the oscillator realization of Lie algebras.
This method has been extended
to the classical (super) Lie algebras
by employing various types of oscillator algebras  such as fermi, para-bose,
and para-fermi algebras as well as Heisenberg-Weyl algebras,
and has been a valuable tool
for the representation theory of Lie algebras\cite{cahn}.

Recently, quantum deformation of a Lie algebra\cite{dri,jim} based on the
deformation of a Lie-Poisson structure, has been actively studied
due to its prominent roles in diverse areas of physics and mathematics
(for a review see \cite{tak} for example).
The deformation has also been studied in various different approaches
such as in the form of a pseudogroup\cite{wor}, as the transformation group of
a non-commutative geometry\cite{manin,wess},
and by using the filtration\cite{rin}.
As in the classical Lie algebra theory,
many works have been done to realize the (super)quantum groups
{\it a la} Schwinger method, by introducing various q-deformed oscillator
algebras starting with the $su_q(2)$ realization in terms of
q-oscillators in \cite{mac,bie,sun}.
In particular, much attention has been paid to the generalized deformed
bose or fermi oscillators of single-mode( \cite{bona1,bona2,melj}
and references therein) or multi-modes\cite{jagan} in itself in connection
with some physical applications as well as with the quantum group.

In this paper, we propose to consider a problem, which is in a sense
an inverse one to the Schwinger method as explained below, and a natural
procedure of deforming the so-obtained algebra. As shown in the text,
various q-deformed algebras are produced together with new ones,
and moreover, many interesting extensions will be possible as commented
in the last section.

Suppose an algebra is given whose generators are factorized into
two independent pairs of operators, with one part taken as a
Heisenberg-Weyl algebra for simplicity.
The problem we address is to determine the other pair of operators
to satisfy the given algebra,
which will be called an inverse problem to the Schwinger realization.
In the next section 2, we will state the problem more precisely
together with our working assumptions
following closely the Schwinger construction of $su(1,1)\,$
algebra as a typical example. The solution is obtained by
solving the defining commutation relations algebraically
considering the full Hilbert space,
which will be referred to as an {\em algebraic} solution.

Next, as a natural means to deform the algebraic solution,
the following procedure is suggested. We take expectation values
on the Hilbert space of the Heisenberg-Weyl algebra
in the defining commutation relations of the given algebra,
and find the operator solution
obeying the resulting {\em effective} commutation relations.
This solution will be called a {\em deformed} solution
and form a deformed algebra of the one
obtained in the inverse Schwinger problem
since it has one more parameter (due to the averaging)
than the algebraic solution which acts like a deformation parameter,
and it reduces to the algebraic one at a special value of the parameter.

In section 3, we illustrate the above ideas with several specific examples.
With a few simplifying assumptions in the first three subsections,
a variety of algebras such as exponential phase operator (q-oscillator),
Heisenberg-Weyl algebra (Calogero-Vasiliev algebra),
and $su(1,1)$ algebra ($su(1,1)$-Calogero-Vasiliev algebra) emerge
as algebraic (deformed) solutions respectively. In the last subsection,
we give an example in which the algebraic solution is absent, while
the deformed one is an interesting q-deformed algebra which becomes
a q-oscillator or a $su_q(1,1)$ algebra depending on the parameters.

In the final section, we give some comments and discuss the extensions
of the method. Some simple formulas related with
the expectation values are given in the Appendix for convenience.

\section{ Inverse Schwinger Method }

We start by recalling the Schwinger construction of $su(1,1)$ algebra
as an example, in terms of two sets of independent oscillators
$\{ \: (a,\ad, n_a), (b, \bd, n_b) \: \}$ which satisfy the
Heisenberg-Weyl algebras as in \eq{oscillator}.
For a given  $su(1,1)$ algebra,
\bea
\com{J_-}{J_+}{2J_0} \: , \nonumber  \\
\com{J_0}{J_\pm}{\pm J_\pm}, \ll{su11}
\eea
the Schwinger method realizes the algebra in the product forms of oscillators
as follows.
\be
J_+ \= \ad\bd \; ,\;\; J_- \= b a \; ,\;\; J_0 \= \trac{1}{2}(n_a \+ n_b \+ 1).
\ll{Jab}
\ee
We note the following property of the Schwinger construction.
\bea
\com{n_i}{J_\pm}{\pm J_\pm}, \nonumber\\
\com{n_i}{J_0}{0}, \;\;\;\;\;\;\;\;\;\; i\= a,b. \ll{nJ}
\eea
It is natural to think that the property \eq{nJ} will determine the Schwinger
construction of $J_\pm$ as in \eq{Jab}, once $J_0$ is chosen as in \eq{Jab}.
This will be an inverse procedure of the Schwinger method,
so we call it the {\em inverse Schwinger method}.

Let's formulate the inverse Schwinger method by taking close analogy with
the above $su(1,1)$ example. A similar work can be done following the
$su(2)$ construction of Schwinger as well. We introduce $\{ D_- , D_+ , \T \}$
instead of $\{ J_- , J_+ , J_0 \}$, and assume
the following commutation relations resembling \eq{su11} and \eq{nJ}:
\bea
\com{D_-}{D_+}{\T}, \ll{DD} \\
\com{n_i}{D_\pm}{\pm D_\pm}, \ll{naD} \\
\com{n_i}{\T}{0}, \;\;\;\;\;\;\;\;\; i\= a,b. \ll{nT}
\eea
It will be natural to take $\T$ as a function of $n_a$ and $n_b$ in view of
the last commutator \eq{nT},
\be
\T \= \T (n_a, n_b). \ll{T}
\ee
We note that the commutation relations between $\T$ and $D_\pm$
need not be specified, since they are obtained by \eq{naD}.
Then the algebra satisfies the Jacobi identity for
the generators $\{ n_a , n_b ,D_+ , D_- \}$.

We briefly consider about the representation of the $D$-operator, which
will be only formal and not be explicit since the $\T$ is not specified yet.
In the Hilbert space $\{ \:  \mid \, m_a, m_b \, \ra \}$, where
$n_i \, \:  \mid \, m_a, m_b \, \ra \=  m_i \, \:  \mid \, m_a, m_b \, \ra ,
\;\; m_i\= 0, 1, 2, \cdot\cdot\cdot ,  \;\;\; i\= a,b $,
the commutator \eq{DD} is reduced to the following difference equation :
\be
F( m_a + 1 , m_b + 1 ) - F( m_a , m_b ) \=  \T ( m_a , m_b ), \ll{de}
\ee
where $F( m_a , m_b )$ is the matrix element of $D_+D_-$.
By noting that $n_a - n_b$ commutes with all the generators
$\{ n_a , n_b ,D_+ , D_- \}$, \eq{de} can also be considered as a difference
equation for a function of $l$ and $s$,
where $m_a = l + s$ and  $m_b = l - s$. The inverse problem is equivalent to
solving \eq{de} for a given $\T$
with appropriate initial conditions on $F( m_a , m_b )$.

{}From now on, we consider  $D_\pm$ operators in the product form of
$a$- and $b$-contribution. In order not to deviate too large from
the Schwinger construction, we choose one part of $D_\pm$ as an
Heisenberg-Weyl generators $ (a, \ad) $, and the other part,
denoted as $B_\pm$ which are to be determined by
\eq{DD} or \eq{de} for a given $\T$,
is assumed to commute with $ (a, \ad) $. That is, we put
\be
D_+ \= \ad B_+ \; , \;\;\;\; D_- \= a B_- \; , \ll{aB}
\ee
and $B_\pm$ satisfy
\bea
\com{a}{B_\pm}{0}\;, \;\;\;\;\;\;\;\;
\tcom{\ad}{B_\pm}{0}\;, \ll{nB} \\
\com{n_b}{B_\pm}{\pm B_\pm}\;, \ll{nBstep}
\eea
where the last commutator  follows from \eq{naD}.

Now, from \eq{DD} and \eq{aB}, we get
\be
a\ad \: B_- B_+ \- \ad a \: B_+ B_- \= \T (n_a,n_b). \ll{BBT}
\ee
Therefore, within our assumptions, $\T$ should be at most the first order
in $n_a$ to yield consistent algebraic solutions for $B$-operators,
\be
\T (n_a,n_b) \= n_a \; G_1 (n_b) \;+\; G_2 (n_b) . \ll{theta}
\ee
Due to \eq{nBstep}, we may put
\bea
B_+ \mid \,m_b \, \ra &=& \sqrt{C ( m_b + 1)} \mid \,m_b + 1 \, \ra ,
\nonumber\\
B_- \mid \,m_b + 1 \, \ra &=& \sqrt{C ( m_b + 1)} \mid \,m_b \, \ra ,
\;\;\;\;\;\; m_b = 0, 1, 2, \cdot\cdot\cdot\;\,.  \ll{Bstep}
\eea
In addition we choose $\mid \,0_b \, \ra$ as the vacuum for $B$-operators
\be
B_- \; \mid \,0_b \, \ra \= 0 \,. \ll{Bvac}
\ee
Then \eq{BBT}$-$\eq{Bvac} lead to a difference equation,
\bea
C ( m_b + 1 ) - C ( m_b ) &=&  G_1 ( m_b )\,, \nonumber \\
C ( m_b + 1) &=& G_2 ( m_b )\,,
 \;\;\;\;\;\; m_b = 1, 2, \cdot\cdot\cdot\;\,,  \nonumber \\
C ( 1 ) &=& G_1 ( 0 ) \=\, G_2 ( 0 ) \;\,.  \ll{Bdiff}
\eea
This is the equation to solve for the inverse Schwinger method, and its
solution will be called an {\em algebraic} solution.
Note that this equation appears also in the investigation of deformed
single-mode oscillators in general (see for example, \cite{melj} ),
although our motivation is different.

As mentioned in the Introduction, we proceed to obtain the deformed algebra
of the above one given by \eq{Bdiff}.
We simply take expectation values on the Hilbert space of the $a$-oscillators
in \eq{BBT} using \eq{aad} and  \eq{ada},
\be
q^2 \; ( \tB_- \tB_+  - G_2 ( n_b ) ) -
( \tB_+ \tB_- + G_1 ( n_b ) - G_2 ( n_b )) \=  0 \,, \ll{Qdiff}
\ee
where $q^2 \= e^{\b \ep_a}$, $\ep_a$ is the quantum level
of the $a$-oscillator, and $\;\; \tilde{} \;\;$ is used to denote that
the expectation values are taken for the $a$-oscillators.
Referring to \eq{nBstep}, it still holds that
\bea
\com{n_b}{\tB_\pm}{\pm \tB_\pm} \;. \ll{tBncomm}
\eea
Thus putting as in \eq{Bstep},
\bea
\tB_+ \mid \,m_b \, \ra &=& \sqrt{\tc ( m_b + 1)} \mid \,m_b + 1 \, \ra ,
\nonumber\\
\tB_- \mid \,m_b + 1 \, \ra &=& \sqrt{\tc ( m_b + 1)} \mid \,m_b \, \ra ,
\;\;\;\;\;\; m_b = 0, 1, 2, \cdot\cdot\cdot\;\,,  \ll{tBstep}
\eea
and with the choice of vacuum as in \eq{Bvac},
\be
\tB_- \, \mid \,0_b \, \ra \= 0 \,, \ll{tBvac}
\ee
we get from \eq{Qdiff} the following equation for the deformed solution :
\bea
q^2 \; ( \tc( m_b + 1 ) - G_2 ( m_b ) ) -
( \tc( m_b ) + G_1 ( m_b ) - G_2 ( m_b )) &=&  0 \,,
 \;\;\;\;\;\; m_b = 1, 2, \cdot\cdot\cdot\;\,, \nonumber \\
q^2 \; ( \tc( 1 ) - G_2 ( 0 ) ) \=  G_1 ( 0 ) - G_2 ( 0 ) \,. \ll{qdiff}
\eea

Now, we give some comments.
Taking expectation values can be viewed as a thermal
average physically if we imagine a system
whose hamiltonian is $h_a = \ep_a \, n_a$
and the $B$-operators are distinguishing the degenerate energy eigenstates
like in a Landau problem for instance.  Alternatively, any relevant
weighted average may be taken associated with the system under
consideration described by commuting  operators $a_\pm$ and $B_\pm$.
Compared with  \eq{Bdiff} which fully takes account of the operator equation
\eq{BBT}, the above \eq{qdiff} may contain only
limited, or in a sense averaged, informations connected to the $a$-oscillator.
Thus, we call \eq{Qdiff} or \eq{qdiff} as an {\em effective}
commutation relation of \eq{BBT}. Since the solution to \eq{qdiff}
will contain the parameter $q$ and reduce to the algebraic solution to
\eq{BBT} or \eq{Bdiff} as $q \r  \infty$,
it will be referred to as a {\em deformed}
solution with a deformation parameter $q$. In the next section, we examine
simple cases of the inverse Schwinger method and present the corresponding
algebraic and deformed solutions as illustrations of using it to obtain
deformed algebras.

\section{Algebraic Solutions and Deformations}

We restrict our attention to the following forms of $G_1 ( m_b )\,$,
$G_2 ( m_b )\,$,
\bea
G_1 ( m_b ) &=& g ( m_b + 1) - g ( m_b ) \,, \nonumber \\
G_2 ( m_b ) &=& g ( m_b + 1) \,,
\;\;\;\;\;\;\;\;\;\; m_b = 0, 1, 2, \cdot\cdot\cdot\;\,, \ll{dform}
\eea
in which case the difference equations \eq{Bdiff}\,, \eq{qdiff} become
{\em homogeneous} ones. An example for the inhomogeneous difference
equation will be given in subsection 3.4 .

To get the algebraic solution, we should consider $g ( 0 )\,= 0\,$
in solving \eq{Bdiff} in order to satisfy the initial condition
in \eq{Bdiff}.
Then the algebraic solution is given by
\be
B_-\,B_+ \= g ( n_b + 1)\,,\;\;\; B_+\,B_- \= g ( n_b )\,. \ll{SS}
\ee

The deformed solution is obtained from \eq{qdiff} as
\bea
\tB_-\,\tB_+ &=& g ( n_b + 1) \- g_0\;q^{-2(n_b+1)}\,,  \nonumber \\
\tB_+\,\tB_- &=& g ( n_b ) \- g_0\;q^{-2n_b}\,, \ll{WS}
\eea
where $g_0 \= g ( 0 )\,$ is not restricted to zero in this case.

To avoid the possible misunderstandings, we stress that the solution
to \eq{Bdiff}, which we are referring to as the algebraic solution
\eq{SS}, is the true solution to the problem
we termed as an inverse Schwinger problem
and exists only if $g ( 0 ) \= 0\,$ for the choice of vacuum as in
\eq{Bvac}. However, the effective commutation relation \eq{Qdiff},
which is obtained by taking expectation values on the $a$-oscillators
in \eq{Bdiff}, admits a more general solution $\tB_\pm$
given in \eq{WS} sharing the same properties as $B_\pm$
(compare equations \eq{tBncomm}, \eq{tBstep}, \eq{tBvac}
with \eq{nBstep}, \eq{Bstep}, \eq{Bvac}). In addition
$\tB_\pm$ are reduced to  $B_\pm$ as  $q \r  \infty$,
which suggests us to interpret them as deformed operators of
$B_\pm$ with a deformation parameter $q$.

In the following subsections 3.1$-$3.3, we examine some examples of solutions
in detail with $g ( m_b )\,$ of the form,
\be
g ( m_b ) \= c_0 + c_1\,m_b + c_2\,m_b^2\,,
\;\;\;\;\;\; m_b = 1, 2, \cdot\cdot\cdot\;\,, \ll{Pform}
\ee
taking $g ( 0 )\,$ to be zero for the algebraic solution
or as a non-zero parameter for the deformed solution respectively
as mentioned above.
This must be the most simple generalization
of the Schwinger realization of $su(1,1)$ case
which corresponds to $\,c_0 = c_2 = 0\,,\; c_1 = 1\,$, and
$g ( 0 )\,= 0\,.$

\subsection{Exponential phase operator and its q-deformation}

We consider the case  $c_1 = c_2 = 0\,$
and normalize $c_0 = 1\,$ in \eq{Pform}. In this case, \eq{SS} is
\bea
B_-\,B_+ &=& 1 \,, \nonumber \\
B_+\,B_- &=& 1\- \mid 0_b \ra \la 0_b \mid  \,. \ll{nonunitary}
\eea
The explicit forms of $B_\pm\,$ are given as (see \eq{Bstep})
\bea
B_- &=& \sum_{m_b =0}^{\infty} \: \mid m_b \ra \la m_b \+ 1 \mid
\: \equiv \: e^{i\f}\:, \nonumber\\
B_+ &=& \sum_{m_b =0}^{\infty} \: \mid m_b \+ 1 \ra \la m_b \mid
\: \equiv \: e^{-i\f}\:.\ll{expop}
\eea
These have been known as Susskind and Glogower's exponential phase operators
\cite{suss,carr} which are non-unitary and non-commuting.
Equations \eq{nBstep} and \eq{Bvac} are
\bea
\com{n_b}{e^{\mp i\f}}{\pm \: e^{\mp i \f}}\: ,\\
e^{i\f} \: \mid 0_b \ra &=& 0 \: .
\eea
We note that the $b$-oscillators obeying the Heisenberg-Weyl algebra are
expressed in the `polar' decomposition form as following
by using the above exponential phase operators :
\be
b \= e^{i\f}\:\sqrt{n_b} \: , \;\;\;\;\;\;
\bd \= \sqrt{n_b} \: e^{-i\f} \: .\ll{polar}
\ee

Let's consider the deformed solution \eq{WS}
given as
\bea
\tB_-\,\tB_+ &=& 1 \- g_0\;q^{-2(n_b+1)}\,,  \nonumber \\
\tB_+\,\tB_- &=& (\: 1 \- g_0\;q^{-2n_b}) \,(\: 1\- \mid 0_b \ra \la 0_b \mid
\:)\,. \ll{BpBm}
\eea
If we take $g_ 0 = 0\,$, \eq{nonunitary} is recovered.
Thus a non-zero $g_ 0\,$ gives the q-deformation.
Equations \eq{tBncomm} and \eq{Qdiff} are
\bea
\com{n_b}{\tB_\pm}{\pm \tB_\pm}\:, \nonumber \\
q^2 \;  \tB_- \tB_+  -  \tB_+ \tB_-  &=&  q^2 \- 1
+ (\: 1 \- g_0\:) \mid 0_b \ra \la 0_b \mid \:, \ll{qoscillator}
\eea
and we recall that the vacuum is chosen as usual in \eq{tBvac}.
Note that when $g_0 = 1\,$, \eq{qoscillator} is a version of
the q-deformed oscillator introduced in \cite{mac}.

Introducing the q-deformed number operator $\qnb$ as
\be
\qnb \= 1 \- g_0\: q^{-2n_b} \,,
\ee
we can write the explicit form of $\tB_\pm\,$ in terms of the above
exponential phase operator $B_\pm\,$
\bea
\tB_- &=& e^{i\f} \: \sqrt{\qnb}\:,\nonumber \\
\tB_+ &=& \sqrt{\qnb} \: e^{-i\f}\:. \ll{Bpolar}
\eea
This form is essentially equivalent to that of $b$-oscillator \eq{polar}
with changes only in the `amplitude' part
by replacing the normal number operator of $b$-oscillator
into the q-number operator $\qnb\,$.
Thus the Hilbert space for the $\tB$-operators is the same as
that of the exponential phase operator.
Since the $\tB_\pm\,$ reduce to the exponential phase operator
as $q \r  \infty\,$, they may be considered as a $q$-deformation of the
exponential phase operators.

It is clear from above that q-deformation alters the amplitude part of
the operator leaving the phase part, the exponential phase operator
unchanged, which is a well-known fact.
This feature shows up repeatedly in the following subsections
as in the second equations of \eq{nonunitary}, \eq{BpBm}, \eq{qoscillator}.
It does not cause any difficulty in writing a polar decomposition form
for a deformed operator as seen in \eq{Bpolar}. However, we will
frequently pay attention only to a case where
the commutation relation takes a simple form ( {\it e.g.}
take $g_0 = 1\,$ in \eq{qoscillator} ).

\subsection{Oscillator and Calogero-Vasiliev oscillator}

Let's consider the case where  $c_2\= 0\:,
\: c_1 \= 1 \:$ in \eq{Pform}.
The algebraic solution \eq{SS} is
\bea
B_-\,B_+ &=& n_b \+ 1 \+ c_0 \,, \nonumber \\
B_+\,B_- &=& (\:n_b \+ c_0 \: )\, (\:1 \- \mid 0_b \ra \la 0_b \mid \:)\,.
\eea
Using the exponential phase operator \eq{expop}
introduced in the previous subsection,
$B_\pm\,$ are
\be
B_- \= e^{i\f}\: \sqrt{c_0\+ n_b} \:, \;\;\;\;\;\;
B_+ \= \sqrt{c_0\+ n_b} \: e^{-i\f}\:.
\ee
When $c_0 = 0\,$, these $B_\pm$ become  normal $b$-oscillators
and correspond to the Schwinger realization of $su(1,1)\,$.

The deformed solution \eq{WS} is given as
\bea
\tB_-\,\tB_+ &=& n_b \+ 1 \+ c_0 \- g_0\, q^{-2(n_b+1)}\,, \nonumber \\
\tB_+\,\tB_- &=& (\: n_b \+ c_0 \- g_0\, q^{-2n_b} \:)\;
(\:1 \- \mid 0_b \ra \la 0_b \mid \:)\,.
\eea
The $\tB_\pm\,$ can also be written in the polar form as in section 3.1
with changes occurring in the amplitude part only.

Considering the case $c_0  = g_0\,$( see the remarks at the end of section 3.1
),
the commutator for  $\tB_\pm$ is given by
\be
\tcom{\tB_-}{\tB_+}{ 1\- c_0 (q^{-2} \- 1) \:  q^{-2n_b} } \ll{suq11}
\ee
with $\tcom{n_b}{\tB_\pm}{\pm \: \tB_\pm}$.
This algebra is a q-deformation of a usual Heisenberg-Weyl algebra
which is recovered  as $q \r  \infty\,$.

Let us note that if we put formally $q^{-2}\,= - 1\,$, the algebra \eq{suq11}
is the Calogero-Vasiliev\footnote{This is the simplest case of
the extended Heisenberg algebra involving exchange operators
\cite{brink,poly,dunkel} which is widely used in the study of the integrable
Calogero-Moser-Sutherland type many-body systems. It is also equivalent to
the single-level para-bose algebra\cite{onu,muku,mac2}
and its q-deformation is performed in \cite{mac3,rim}.} algebra :
We change the notation $\tB_-\: ,\: \tB_+ \:,\: c_0\,$
into $ C \: , \: C^\dagger \:,\: \nu \,$ respectively. Then we have
\bea
C\,C^\dagger &=& n_b \+ 1 \+ \nu \+ \nu \,K \,, \nonumber \\
C^\dagger\,C &=& n_b \+ \nu \- \nu \,K\,,\ll{Calo}
\eea
where the parity operator $K$ is given as
\bea
K &=&  (-1)^{n_b}\,, \;\;\;\;\;\; K^2 \= \: 1 \,,\nonumber \\
K\: C &=&  -\: C\: K\,,\;\;\;\;\;\; K\: C^\dagger \= -\: C^\dagger\:K \,.
\ll{Parity}
\eea
The explicit forms of $ C \: , \: C^\dagger \:,$  are given as
\be
C \= e^{i\phi}\:\sqrt{ n_b \+ \nu \- \nu\,K }\,,\;\;\;\;\;\;
C^\dagger \= \sqrt{ n_b \+ \nu \- \nu\,K }\:e^{-i\phi}\,. \ll{CVOP}
\ee
Thus, Calogero-Vasiliev algebra with the deforming parameter $\nu\, $
is obtained
\bea
\com{C}{C^\dagger}{1\+ 2\nu K}\,,\nonumber \\
\com{n_c}{C}{-C}\,,\;\;\;\;\;\; \tcom{n_c}{C^\dagger}{C^\dagger}\,,\
\ll{calogero}
\eea
where the number operator $n_c\,$ is defined as
\be
n_c \= \trac{1}{2}\,( \:C\,C^\dagger \+ C^\dagger\,C \- 1 \:)
    \= \, n_b \+ \nu \;.
\ll{CVn}
\ee

\subsection{Holstein-Primakoff and Calogero-Vasiliev extension of $su(1,1)$}

Let's consider the algebraic solution  \eq{SS} for $c_2 \= 1\,.$
In this case the solution is
\bea
B_- \, B_+ & = & ( n_b \+ 1 ) ( n_b \+ 1 \+  c_1 ) \+ c_0 \,,\nonumber \\
B_+ \, B_- & = & (\:n_b ( n_b \+  c_1) \+ c_0\:)\:
 (\:1 \- \mid 0_b \ra \la 0_b \mid \:)\,.
\eea
Specializing to the case $c_0 = 0\,$, $B_\pm\,$ together with $B_0\,$
defined as
\be
B_0 \= n_b \+ \trac{1}{2}( c_1 \+ 1 )\,,
\ee
realize the $su(1,1)$ algebra
known as the Holstein-Primakoff realization\cite{hol},
\bea
B_- &=& b\: \sqrt{ B_0 \+ \trac{1}{2}( c_1 \- 1 ) }\,, \nonumber\\
B_+ &=& \sqrt{ B_0 \+ \trac{1}{2}( c_1 \- 1 ) } \: b^\dagger \,, \nonumber\\
\com{B_-}{B_+}{2\,B_0} \,,\;\;\;\;\;\;
\tcom{B_0}{B_\pm}{\pm\,B_\pm}
\,.\ll{hpreal}
\eea

The deformed solution \eq{WS} is obtained as
\bea
\tB_-\,\tB_+ &=& ( n_b \+ 1 ) ( n_b \+ 1 \+  c_1 ) \+ c_0
\- g_0\,q^{-2(n_b+1)} \,, \nonumber \\
\tB_+\,\tB_- &=& (\:n_b ( n_b \+  c_1) \+ c_0\:
\- g_0\,q^{-2n_b} \:) \:(\:1 \- \mid 0_b \ra \la 0_b \mid \:)\;.
\eea
Let us consider the case $c_0 = g_0\,$.
Then $\tB_\pm\,$ together with $\tB_0\,$ defined as
\be
\tB_0 \= n_b \+ \trac{1}{2}\,(\: c_1 \+ 1 \+ c_0\,( 1 \- q^{-2} )
\,q^{-2n_b}\:)\;,
\ee
realize a  q-deformed $su(1,1)$ algebra (not in a standard form),
\be
\tcom{\tB_-}{\tB_+}{2\,\tB_0}\;, \;\;\;\;\;\;
\tcom{B_0}{\tB_\pm}{\pm\,\tB_\pm}\;,
\ee
and reduce to the above
Holstein-Primakoff realization \eq{hpreal} as $q \r  \infty\,$.

In this case, if we put formally $q^{-2}\, = -\,1$
as in the previous subsection, we  find an interesting deformed
$su(1,1)$ algebra realized by the Calogero-Vasiliev oscillator
$C, C^\dagger\,$ introduced in \eq{Calo}$-$\eq{CVn} :  Let $c_1\= 2c_0 + 1\,$.
We change notations of $\tB_\pm\:, B_0\;, c_0 \,,$ into
$J^c_\pm \;,J^c_0 \;, \nu \,,$ respectively. Their explicit forms are
\bea
J^c_- &=& C\: \sqrt{ n_c \+ 1 \+ \nu\, K } \,, \nonumber \\
J^c_+ &=& \: \sqrt{ n_c \+ 1 \+ \nu\, K } \: C^\dagger \,, \nonumber \\
J^c_0 &=& \: n_c \+ 1 \,,
\eea
where the parity operator $K\,$ and the number operator $n_c\,$ of the
Calogero-Vasiliev oscillator are given in equations \eq{Parity},
\eq{CVn}. The afore-mentioned deformed $su(1,1)$ is obtained as
\bea
\com{J^c_-}{J^c_+}{2 \,J^c_0 \+ 2\, \nu \, K }\;,\;\;\;\;\;\;
\tcom{J^c_0}{J^c_\pm}{\pm\, J^c_\pm} \;,\nonumber \\
J^c_\pm \,K &=& - K \,J^c_\pm ,\;\;\;\;\;\; J^c_0 \,K \=  K \,J^c_0 \;,
\eea
which reduce to the Holstein-Primakoff realization \eq{hpreal}
of $su(1,1)$ algebra by an ordinary oscillator when $\,\nu = 0\,$.

\subsection{q-Holstein-Primakoff realization of $su_q(1,1)$}

As an example of more complicated case of the inverse Schwinger method,
let's consider $\T$ which leads to an
inhomogeneous difference equation from \eq{BBT}.
We take $\T$ as
\be
\T (n_a,n_b) \= \t ((n_a + \trac{1}{2} -\lambda)(\mu + \trac{1}{2} -n_a)) \: ,
\ll{finite}
\ee
where $\t (x)$ is the Heaviside step function
and $\mu\,, \lambda\,$ are integers of $\,0 \le \lambda \le \mu$.
The function takes unity as its value in the finite interval
$\lambda \le n_a \le \mu$
with $(\mu - \lambda + 1)$-independent states.

The equation \eq{BBT} has no solution for above $\T$, however
on taking expectation values on the $a$-oscillator,
it admits a nontrivial solution.
Using \eq{step}, the expectation value of the step function \eq{finite}
is given as
\be
\la \, \t ((n_a + \trac{1}{2} -\lambda)(\mu + \trac{1}{2} -n_a)) \, \ra
\= q^{-2\lambda} ( 1 \- q^{-2(\mu - \lambda + 1 )} )\,,
\ee
and the resulting effective commutation relation is
\be
q^2 \: \tB_-\,\tB_+ \- \tB_+\,\tB_- \= (q^2 \- 1) \: q^{-2\lambda}
( 1 \- q^{-2(\mu - \lambda + 1 )} )\,,\ll{Q}
\ee
and $\tcom{n_b}{\tB_\pm}{\pm \: \tB_\pm}$ from \eq{tBncomm}.

With the vacuum as in \eq{tBvac}, the solution to \eq{Q} is easily
obtained and related with the q-deformed operators given in section 3.1
by a simple normalization.
When we take
\be
\mu \; \rightarrow \; \infty \: ,\;\;\;\; \lambda \= n_b \: ,
\ee
the solution to \eq{Q} is also related  to the q-deformed operators
of section 3.1 by a $n_b$-dependent transformation.

Another interesting case is
\be
\mu \= 2 (n_b \+ \sigma) \- 1 \: ,\;\;\;\;\;\; \lambda \= 0 \: .
\ee
Transforming $\{\tB_\pm \;, n_b \}$ into new generators
$\{ \tJ_\pm\; , J_0 \}$ related by
\bea
\tJ_- &=& (q^2 \- 1)^{-1} \: \tB_- \, q^{n_b + \sigma + 1/2} \: ,\nonumber \\
\tJ_+ &=& (q^2 \- 1)^{-1} \: q^{n_b + \sigma + 1/2} \, \tB_+ \: ,\nonumber \\
J_0 &=& n_b \+ \sigma \: ,
\eea
then $\{ \tJ_\pm\; , J_0 \}$ satisfy the $su_q(1,1)$ algebra :
\be
\tcom{\tJ_-}{\tJ_+}{[2 J_0]}\,, \;\;\;\;\;\;
\tcom{J_0}{\tJ_\pm}{\pm \tJ_\pm} \,,
\ee
where $[x] \= (q^x \- q^{-x}) / (q \- q^{-1}) $ is a normal q-number.
Explicitly,
\bea
\tJ_- &=& e^{i\phi}\; \sqrt{\; [n_b]\,[ n_b + 2\,\sigma - 1 ]}
\= b_q \; \sqrt{\;[ n_b + 2\,\sigma - 1 ]}\;,\nonumber \\
\tJ_+ &=& \sqrt{\; [n_b]\,[ n_b + 2\,\sigma - 1 ]}\;e^{-i\phi}
\= \sqrt{\;[ n_b + 2\,\sigma - 1 ]}\; b_q ^\dagger \;,
\eea
with the standard q-oscillators $b_q\;,b_q ^\dagger\,$,
give the q-Holstein-Primakoff realization of $su_q(1,1)$
in terms of a q-oscillator\cite{chaic,kun}.

\section{Discussion}

We have considered an inverse problem to the well-known Schwinger
realization of Lie algebras. Taking Schwinger's $su(1,1)$ construction by two
commuting oscillators as our model case, we have made a few assumptions
to explain the ideas in a simple setting as given in sections 2 and 3.
However, various interesting examples such as the exponential phase
operators, normal oscillators, and the Holstein-Primakoff realization of
$su(1,1)$ algebra appear as solutions to the inverse problem we considered,
although they have been known from other motivations.

More importantly, we have obtained the corresponding q-deformed algebras
naturally by solving the effective commutation relations.
Therefore, on the basis of simple but nontrivial examples
examined in section 3, we may regard the procedure taken in this paper
as an efficient way of obtaining deformed algebras.
It will be worthwhile to understand the mathematical structure
involved in this procedure. In particular, the q-Heisenberg-Weyl algebra
of the subsection 3.1 has been explained on the basis of the
contact metric structure of the Heisenberg-Weyl group manifold\cite{park1}.

If we extend the above inverse Schwinger method to the
fermionic algebra\cite{park2} as in the sections 2 and 3,
we can easily find the various fermionic q-deformed algebras.
Extension to the multi-mode case is also possible.
As a simplest case, one may find a set of independent oscillators
and corresponding q-oscillators.
More nontrivial q-deformed system, such as the $SU_q(N)$-covariant system of
q-oscillators, will appear through similar considerations as
in  the subsection 3.4.
Further detailed description is in preparation,
and will be reported.

\subsection*{Acknowledgements}
The work of K H Cho is supported in part by KOSEF No. 931-0200-030-2,
by SRC program through Seoul National University, and
by the Basic Science Research Institute Program No. 94-2434.
S U Park is supported by the research fund of Jeonju University in 1994.

\section*{Appendix}
\setcounter{equation}{0}
\makeatletter
\@addtoreset{equation}{A}
\rnc{\theequation}{A.\arabic{equation}}
\makeatother
We consider a single-mode oscillator with generators $\{ n_a,\ad, a \}$
obeying the following familiar commutation relations ;
\be
[ \: a \: , \: a^\dagger \: ] \; = \; 1 \, , \;\;\;\;
[ \: n_a \: , \: a \: ] \; = \- \; a \,, \;\;\;\;
[ \: n_a \: , \: \ad \: ] \; = \; \ad \,,\label{oscillator}
\ee
where $n_a \= \; \ad\,a\;$.
The Hilbert space for the oscillator
is spanned by the number eigenstates
$\mid  m_a \rangle ,\;\; m_a  \= 0,1,2,\cdots \,$.
We take the hamiltonian $h_a$
\be
h_a \= \ep_a n_a \= \ep_a \ad a \;,
\ee
where $\ep_a$ is its quantal energy level.
The expectation value of an operator $\Phi$ is defined  by
\be
\la \: \Phi \: \ra \; \equiv \; \frac{\tr e^{-\b h_a} \: \Phi }
{\tr e^{-\b h_a}}\;,
\ee
where the denominator represents the character of the oscillator algebra with
$q^2 \= e^{\b \ep_a}\,$,
\be
\c \= \tr e^{-\b h_a} \= \frac{1}{1 - q^{-2}} \;. \ll{character}
\ee

We present some expectation values for later use :
\begin{eqnarray}
\langle \, \theta(n_a+1/2-\alpha)\,\rangle &=& e^{-\alpha \beta \epsilon_a}
   \= q^{-2\alpha},\ll{step} \\
\langle \, a a^\dagger \, \rangle &=& \frac{1}{1 \, - \, e^{-\beta \epsilon_a}}
  \= \frac{1}{1 - q^{-2}} \;,   \ll{aad} \\
\langle \, a^\dagger a \,\rangle &=&
   \frac{e^{-\beta \epsilon_a}}{1 \, - \, e^{-\beta \epsilon_a}}
   \= \frac{q^{-2}}{1 - q^{-2}} \;,\ll{ada}
\end{eqnarray}
where $\theta(x)$ is a step function.
In \eq{step}, when $\alpha \= 0\,$,
the expectation value of $\t  (n_a + \trac{1}{2})$
is equivalent to that of the unit.

\rnc{\Large}{\normalsize}

\end{document}